# On the question of metallicity and ferromagnetism in charge ordered insulating $Nd_{0.5}Sr_{0.5}MnO_3$ nanoparticles

S. Kundu[a], T. K. Nath[a,*] A. K. Nigam[b], T. Maitra[c] and A. Taraphder[a,d,e]

[a] Department of Physics and Meteorology, Indian Institute of Technology, Kharagpur 721302, India
[b] Tata Institute of Fundamental Research, Homi Bhaba Road, Colaba, Mumbai 400005, India
[c] Department of Physics, Indian Institute of Technology, Roorkee 247667 India
[d] Centre for Theoretical Studies, Indian Institute of Technology, Kharagpur 721302, India
[e] Max Planck Institute fur Physik Komplexer Systeme, Notnitzer str. 38, 01187 Dresden, Germany

## *Abstract*

*The fascinating phenomenon of destabilization of charge/orbital order in $Nd_{0.5}Sr_{0.5}MnO_3$ with the reduction of grain size is critically investigated. Based on our magnetic and transport experiments followed by a theoretical analysis, we analyze various possible mechanisms and try to delineate a universal scenario behind this phenomenon. We, revisit this issue and discuss the overwhelming evidence from experiments on nano and bulk manganites as well as the absence of correlation between size reduction and pressure effects on manganites. We propose a phenomenological understanding on the basis of enhanced surface disorder to explain the appearance of weak ferromagnetism and metallicity in nanoparticles of $Nd_{0.5}Sr_{0.5}MnO_3$. We also provide supportive evidence from an ab-initio electron structure calculation and a recent numerical simulation and argue that the mechanism is universal in all nanosize charge ordered manganites.*

*Corresponding author: tnath@phy.iitkgp.ernet.in
Tel: +91-3222-283862
Area code: 721302,
INDIA

A major surprise in the physics of manganites is that the fairly strong charge and orbital order in many of them is often destabilized by small external perturbations like magnetic or electric field and hydrostatic pressure [1,2]. Quite interestingly, the reduction of grain size has recently been found to have similar destabilizing effects [3-6] inviting obvious connections between the two [4,6]. However, such inferences, although not implausible, have not been founded on any direct experimental observation and are often, at best, speculative. It is, therefore, crucial to seek either direct experimental confirmation for the underlying mechanism, or look for indirect evidence pointing to possible alternatives through a careful analysis and comparison of experimental data against existing information. Two possible scenarios have been envisaged so far in the literature: (i) The reduction of size introduces *an effective hydrostatic pressure modifying the underlying charge and orbital order in the entire grain* through enhanced bulk strain [4,6] and (ii) the entire effect is primarily driven by *surface modifications of the underlying electronic interactions and electronic structure* at the nano-grains with *increased grain surface area coupled with* the proliferation of disorder and defects in the nano-material [7,8]. The modified microscopic interactions and electronic structures at the grain surface lead to a partial reversal of the magnetic and orbital states at the surface leading to the observed changes in the transport properties in the CMR nano systems. Unfortunately, in either case, it has not been possible to link the experimental observations to the proposed scenarios through direct experiments.

Interpretation of the transport and magnetic properties in the nanosize regime in terms of an effective pressure would imply that similar effects must be seen on application of hydrostatic pressure in bulk systems. Under hydrostatic pressure, bulk $La_{0.5}Ca_{0.5}MnO_3$ shows enhanced stability of the charge ordered (CO) state with concomitant increase in the Neel temperature [9] while $Nd_{0.5}Sr_{0.5}MnO_3$ shows a suppression of charge ordering as its resistivity decreases on application of external pressure [1,9]. The two narrow bandwidth systems, $Nd_{0.5}Ca_{0.5}MnO_3$ and $Pr_{0.5}Ca_{0.5}MnO_3$, show non monotonic change (a cross over) in the pressure induced strain[11] and in their transport behavior [12]. On the contrary, *in nanometric form all the above charge ordered systems show similar behavior* – an enhanced ferromagnetism and metallicity with reduction to nanometric size, clearly indicating a suppression of the charge and orbital order [3,5,8,12]. There is, therefore, no one-to-one connection between bulk systems under pressure and reduction of size to nanometric regime. This has also been observed in the x-ray and neutron scattering results of Jirak, *et al.*[8] recently. Their study clearly shows that the lattice distortion, unit cell volume, Mn-O distances and octahedral tilts at room-temperature *Pbnm* perovskite structure of $Pr_{0.5}Ca_{0.5}MnO_3$ and $La_{0.5}Ca_{0.5}MnO_3$ are practically unaffected by the reduction in particle size. All these observations suggest that there are no significant structural reorganizations as the grain size is reduced. In the face of the combined evidence, the simpleminded understanding of the reversal of magnetic and transport properties in the nanometric systems in terms of effective bulk pressures [6] does not seem to work and one has to look for alternatives.

In this context, we study the intermediate bandwidth manganite $Nd_{0.5}Sr_{0.5}MnO_3$ that undergoes a robust CE-type charge ordering (along with orbital order) at around 150 K ($T_{CO}$) with a reduction of crystallographic symmetry from orthorhombic to monoclinic [7]. We look for the effect of grain size reduction from bulk to the nanometric regime (40 nm grain size) and compare our data with existing literature in order to understand the following questions: (i) *Is the enhanced effective hydrostatic pressure in the nanometric regime causing the observed insulator to metal transition* (MIT) by "melting" the charge/orbital order (CO/OO) in the grains?

If so, *does this corroborate the existing literature on the effect of hydrostatic pressure on manganites*, or (ii) does *the CO/OO state melt at the surface of nanometric grains leading to a size-dependent distribution of tunneling gaps across the grains*? The situation envisaged in (ii) could, indeed, lead to an enhanced conduction and finally a metallic state as well.

Based on our detailed magnetic and transport experiments carried out on the samples of $Nd_{0.5}Sr_{0.5}MnO_3$ with different grain size, we critically reanalyze these two different scenarios in order to delineate the possible mechanism. The samples of $Nd_{0.5}Sr_{0.5}MnO_3$ have been synthesized with continuously varying grain size through chemical route, starting from bulk, down to a size of 40 nm. The HRTEM and FESEM images (Fig. 1) show that we have three samples of average grain size a few μm (referred to hereafter as NSMOBULK), 80 nm (NSMO80) and 40 nm (NSMO40), respectively. To investigate possible structural changes due to the size reduction, Rietveld refinement has been carried out using the high resolution x-ray diffraction data of the samples at room temperature (Fig. 1 (a)). All the samples crystallize in orthorhombic structure with Imma space group. From Table 1 it is evident that the lattice parameters vary only slightly with the change of grain size. We also notice that the maximum change in unit cell volume is only about 0.1 % for NSMO40 which translates to an insignificant resultant pressure on the grains, calculated from the Birch-Murnaghan (BM) equation of state [9]. This implies that there is no significant role of pressure in the nanograins. A simple estimate from surface tension (=2S/D, D is the average grain size), assuming spherical grains, would, on the other hand, produce a maximum pressure of about 2.5 GPa (for NSMO40, assuming typical surface tension S = 50N/m for such perovskites). However, this generally overestimates the pressure over the more careful estimates from BM equation. An enhanced surface pressure in the nano size grains requires a reduction of unit cell volume (compared to the bulk). Experimental results, though, do not unequivocally suggest that unit cell volume always decrease with the decrease of grain size; in fact the recent evidence [8,11] and the results from our x-ray data (Table 1) are quite contrary to this assertion.

The zero field-cooled (ZFC) and field-cooled (FC) temperature dependent magnetization data recorded in a Quantum Design MPMS SQUID magnetometer at 500 Oe field are shown in inset (a) of Fig. 2. NSMOBULK sample clearly shows the ferromagnetic (FM) – paramagnetic (PM) transition at around 250 K ($T_C$) and the FM – antiferromagnetic (AFM), CO transition at around 150 K. Strikingly, the CO transition is not detectable in nanosize NSMO40 and NSMO80 samples, clearly showing a destabilization of the antiferromagnetic CO state below 150 K at nanosizes. Most likely, this destabilization predominantly occurs at the magnetically disordered surface of the nano grains showing a net moment even below 150 K, originating, presumably, from the unsaturated superexchange (as well as enhanced double exchange (DE)) interactions at the surface region. These surface modifications are weakened gradually towards the core of the grain, stabilizing the AFM order in the core. A strong bifurcation of the FC and ZFC curves indicates the presence of a high level of anisotropy in the nanometric samples. A large number of crystallographic defects at the grain surface provide pinning sites for the spins, leading to strong magnetic anisotropy. Such a scenario is well supported by the M-H behavior of the samples. The M-H curve at 60 K of NSMO40 demonstrates a ferromagnetic behavior with large coercivity of about 200 Oe and spontaneous magnetization of 1.75 $\mu_B$/f.u. as shown in Fig. 2. Presence of spontaneous magnetization, evident from the positive intercept in the Arrott plot shown in the inset (b) of Fig. 2, confirms incipient ferromagnetic order and indicates that the DE mechanism, primarily responsible for FM states in the manganites, is operative again and begins to dominate over the super-exchange in NSMO40 samples. Moreover, the calculated saturation magnetization of NSMO40 at 60 K (1.9 $\mu_B$ /f.u.) is smaller than that of the NSMOBULK (2.6 $\mu_B$ /f.u., at 200 K) – a clear indication of the partial conversion of AFM to FM order of spins in the nanograins, supporting a core shell type scenario. The M-H behaviors of NSMOBULK at 15 K (AFM like) and at 200 K (FM like) corroborate the already known results on bulk $Nd_{0.5}Sr_{0.5}MnO_3$ (Inset (c) of Fig. 2).

The temperature dependent real part of ac magnetic susceptibility ($\chi_{ac}$) of NSMO40 sample, measured at different frequencies (Fig. 3) shows that the magnitude of the $\chi_{ac}$ is gradually suppressed around the peak region with increase in frequency. This behaviour in $\chi_{ac}$ is usually interpreted to originate from the disordered or short range nature of the ferromagnetism in the sample. Moreover, one can observe the two different slopes around the FM-PM region in $\chi_{ac}$ of NSMO40 sample. A clearer picture of this double slope nature emerges when the derivative of $\chi_{ac}$ is plotted with respect to temperature (inset of Fig. 3). Besides having a minimum, a small shoulder is observed just below this temperature. The shoulder can be related to the Curie temperature corresponding to the surface magnetic order, whereas the minimum corresponds to that of the core of the nanograins. A similar picture of two peaks in temperature dependent resistivity was observed by N. Zhang *et al*. [13] in $La_{0.85}Sr_{0.15}MnO_3$. It is quite expected that there is a broad distribution of the exchange interactions in the disordered and short ranged ferromagnetic grain surface. This is reflected in the initial slow fall of $\chi_{ac}$ (the shoulder in the $d\chi_{ac}/dT$ vs. T plot) with temperature, and the sharp fall thereafter can be related to the phase transition of the comparably more homogeneous core regions of the grains of NSMO40 sample. In comparison, there is no evidence of any double slope structure in the temperature dependent $\chi_{ac}$ (inset of Fig. 3.) of the NSMOBULK sample.

The temperature dependent electronic transport of the samples evolves significantly with the change in grain size (Fig. 4). The insulating nature of NSMOBULK sample is suppressed and a metallic behavior is observed in the NSMO40 sample. The insulating behavior of NSMOBULK is suppressed in a magnetic field of 9 T, due to field induced melting of CO, and the metal insulator transition temperature of NSMO40 is increased from 127 K to 157 K [Fig. 4 and Inset of Fig. 4]. All these features of magnetic and electronic properties in the nanosize samples, which are almost universal across different CO manganites, could be accounted from a phenomenological picture based on the surface disorder in the nanograins. This phenomenological model is pictorially shown in Fig. 5. Refer to our picture of nanograins discussed above, a disordered shell with a gradually ordered core region, with the surface effect or the proportion of shell thickness increases as the size of the grain decreases. We emphasize the fact that there is a gradual change in the magnetic order from the center to the boundary of these grains. At a certain grain size, the ferromagnetic shells of the nearby grains form a conducting path due to sufficient overlap of the electrons via double exchange leading to a gradual onset of a metallic behavior in the NSMO40. The nature of the states in the core region does not interfere in such a transport evidently. Interestingly, NSMO80 sample does not show a metallic behavior, only a suppression of resistivity compared to NSMOBULK below the CO temperature. This is due to a subdued surface effect, having larger grain sizes than NSMO40. Of particular interest is the fact that the $T_{MI}$ for NSMO40 is around 127 K, whereas that for a bulk $Nd_{0.5}Sr_{0.5}MnO_3$ system is generally found near its $T_C$ (250 K). While the transport of NSMO40 sample is entirely driven by the reconstruction of the magnetic order at the surface of the grains in the scenario discussed, the surface ferromagnetism of the nano grains is very weak compared to the bulk FM. In addition, it is also highly disordered and therefore unable to sustain the magnetic order at higher temperatures leading expectedly to a lower $T_{MI}$. Indeed, $T_{MI}$ is actually reduced down to a value of 127 K.

In order to glean an understanding of this electronic reorganization at the surface, we perform an ab-initio calculation for the electronic structure of the bulk and the surface. It is already known that the surface has extra charges at NdO termination. According to Dong et al.,[15] this introduces a small, attractive, nearly unscreened image potential at the surface. Such a situation may enhance oxidation of the free surface and imply presence of extra oxygen atoms at the surface as well [15, 16]. We have performed first principle calculation using plane wave basis as implemented in Vienna ab-initio simulation package (VASP) [17] within generalised gradient approximation (GGA) [18]. Our results indicate (Fig. 6) that the band structures are almost similar in the two cases, albeit with a small enhancement of the surface DOS at the Fermi level. On the basis of the MC simulation on a DE model with electron-phonon coupling, it was found [15] that these extra charges lead to a weak FM tendency at the surface. An enhanced spin flip process in manganite surfaces leading to FM tendencies via stronger DE mechanism was also reported earlier [19]. Clearly, the marginal enhancement of the DOS at the surface (Fig. 6 inset) also indicates an enhanced DE mechanism that would lead to a destabilization of the AFM order at the surface. We emphasize that all these theoretical analyses point to a very slight modification in surface electronic (and magnetic) states and are only indicative at this stage. Much more detailed theoretical and experimental analysis are necessary to actually arrive at a quantitative resolution.

In summary, we have addressed the issue of partial melting of charge and AFM order due to size reduction to nano scale in $Nd_{0.5}Sr_{0.5}MnO_3$. We discussed the possible underlying mechanisms for the phenomenon and presented experimental and theoretical evidences. The destabilization of CO in nanosize $Nd_{0.5}Sr_{0.5}MnO_3$, at this stage, is primarily attributed to the surface reconstruction of the electronic states and the consequent renormalization of magnetic exchange. Although the destabilization of the CO/AFM order at an individual grain surface is small, the hugely enhanced surface to volume ratio in nano-grains add up to the sizable change in magnetic and transport properties observed in experiments. Connectivity between the (nominally metallic) grain surfaces thereafter produces the overall metallic behavior below a certain temperature in the nanosize NSMO40 sample. Our phenomenological hypothesis as well as theoretical analysis is clearly very general and system independent and explains qualitatively, to a large extent, the observed universality of behavior (of enhanced ferromagnetism and metallicity) in the nano size half doped manganites which are AFM insulators in their bulk form.

## Acknowledgement

One of the authors (T. K. Nath) would like to acknowledge the financial assistance of Department of Science of Technology (DST), New Delhi, India through project no. IR/S2/PU-04/2006. We also acknowledge discussions with A. K. Raychaudhuri and I. Das. TM acknowledges Computational Materials Science Department, University of Twente, Netherlands for providing computational facilities.

## References

[1]Y. Moritomo, H. Kuwahara, Y. Tomioka and Y. Tokura, Phys. Rev. B **55**, 7549 (1997).

[2]M. Imada, A. Fujimori and Y. Tokura, Rev. Mod. Phys. **70**, 1039 (1998); *Colossal Magnetoresistive Oxides* edited by Y.Tokura,Gordon and Breach Science, Singapore, 2000.

[3]T. Zhang, C. G. Jin, T. Qian, X. L. Lu, J. M. Bai, and X. G. Li, J. Mater. Chem. **14**, 2787 (2004); A. Biswas, I. Das, and C. Majumdar, J. Appl. Phys. **98**, 124310 (2005); S. S. Rao, K. N. Anuradha, S. Sarangi, and S. V. Bhat, Appl. Phys. Lett. **87**, 182503 (2005).

[4]Tapati Sarkar, P. K. Mukhopadhyay, A. K. Raychaudhuri and S. Banerjee, J. Appl. Phys. **101**, 124307 (2007).

[5]L Liu, S L Yuan1, Z M Tian, X Liu, J H He, P Li, C H Wang, X F Zheng and S Y Yin, J. Phys. D: Appl. Phys. **42,** 045003 (2009).

[6]Tapati Sarkar, A. K. Raychaudhuri, and Tapan Chatterji, Appl. Phys. Lett. **92**, 123104 (2008).

[7]S. Kundu, T. K. Nath, A. K. Nigam and A. Taraphder, in *Recent Advances in Correlated System*s, Guwahati (2010).

[8]Z. Jirák, E. Hadová, O. Kaman, K. Knížek, M. Maryško, and E. Pollert, Phys. Rev. B **81**, 024403 (2010).

[9]D. P. Kozlenko, L. S. Dubrovinsky, I. N. Goncharenko, B. N. Savenko,V. I. Voronin,E. A. Kiselev, and N. V. Proskurnina, Phys. Rev. B**75**, 104408 (2007).

[10]Congwu Cui, Trevor A. Tyson, and Zhiqiang Chen, Phys. Rev. B **68**, 214417 (2003).

[11]Anthony Arulraj, Robert E. Dinnebier, Stefan Carlson, Michael Hanfland, and Sander van Smaalen, Phys. Rev. Lett. **94,** 165504 (2005).

[12]Congwu Cui and Trevor A. Tyson Phys. Rev. B **70**, 094409 (2004).

[13]S S Rao and S V Bhat, J. Phys. D: Appl. Phys. **42,** 075004 (2009) .

[14]Ning Zhang, Weiping Ding, Wei Zhong, Dingyu Xing, and Youwei Du, Phys. Rev. B **56**, 8138 (1997).

[15]Shuai Dong, Rong Yu, Seiji Yunoki, J.-M. Liu, and Elbio Dagotto, Phys. Rev. B **78**, 064414 (2008).

[16]H. Zenia, G. A. Gehring, G. Banach, and W. M. Temmerman, Phys. Rev. B **71**, 024416 (2005).

[17]G. Kresse and J. Hafner, Phys. Rev. B 47, RC558 (1993); Phys. Rev. B **49**, 14251 (1994).

[18]J. P. Perdew, K. Burke, and M. Ernzerhof, Phys. Rev. Lett. **77**, 3865 (1996).

[19]Alessio Filippetti and Warren E. Pickett, Phys. Rev. Lett. **83**, 4184 (1999).

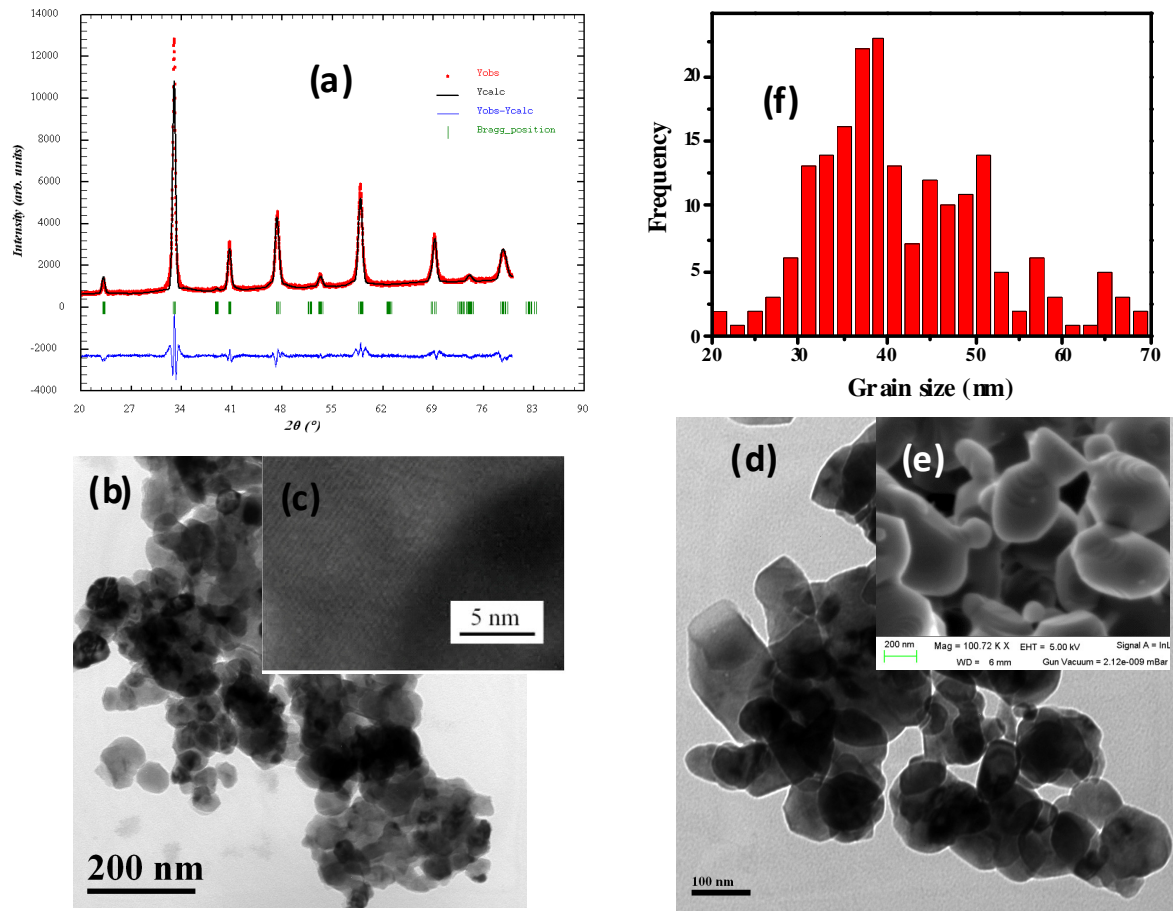

FIG. 1. (a) (Color online) Experimental x-ray diffraction data, fitted curve after Rietveld refinement and difference plot. (b) TEM image and (c) high resolution lattice image of NSMO40 grains. (d) TEM image of NSMO80. (e) FESEM image of NSMOBULK. (f) Distribution of grain size of NSMO40.

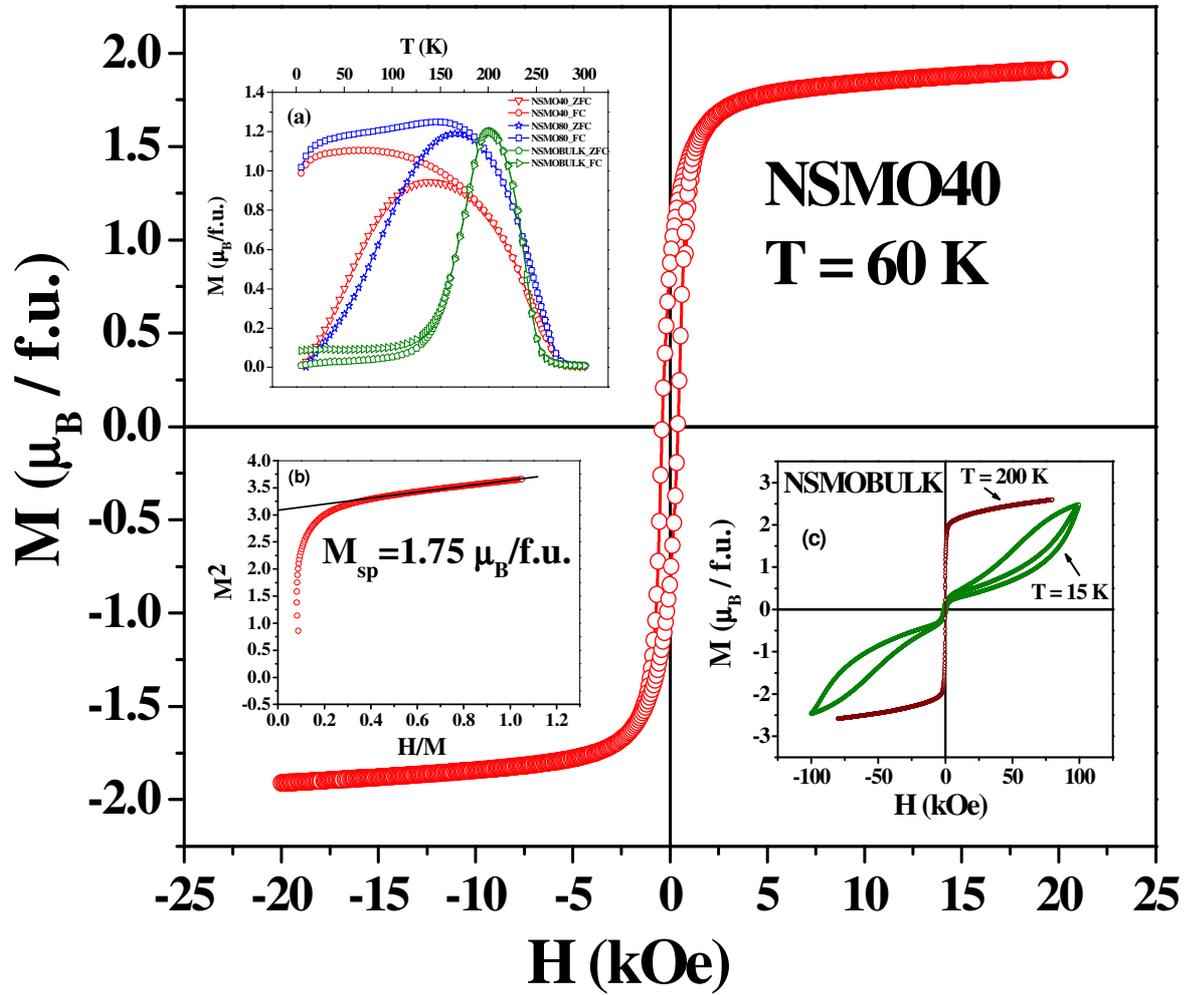

FIG. 2. (Color online) The hysteresis (M vs. H) behaviour of NSMO40 sample at 60 K. Insets: (a) Plot of the FC-ZFC magnetization for all the samples. (b) Arrott plot at 60 K for NSMO40. (c) M-H behaviour of NSMOBULK at 15 K and 200 K.

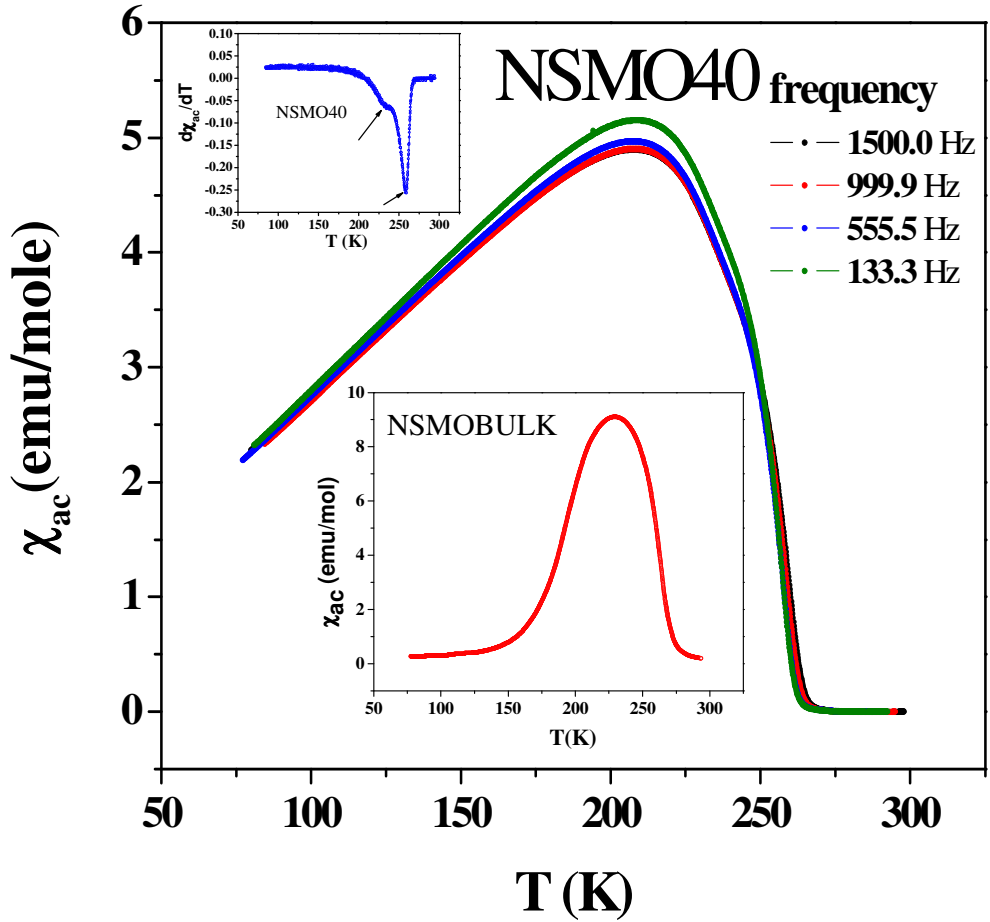

FIG. 3. (Color online) Variation of real part of ac susceptibility ($\chi_{ac}$) of NSMO40 with temperature at different frequencies. Top left inset shows the derivative of $\chi_{ac}$ at the highest frequency. The bottom inset shows real part of $\chi_{ac}$ of NSMOBULK as a function of temperature recorded at 555.5 Hz.

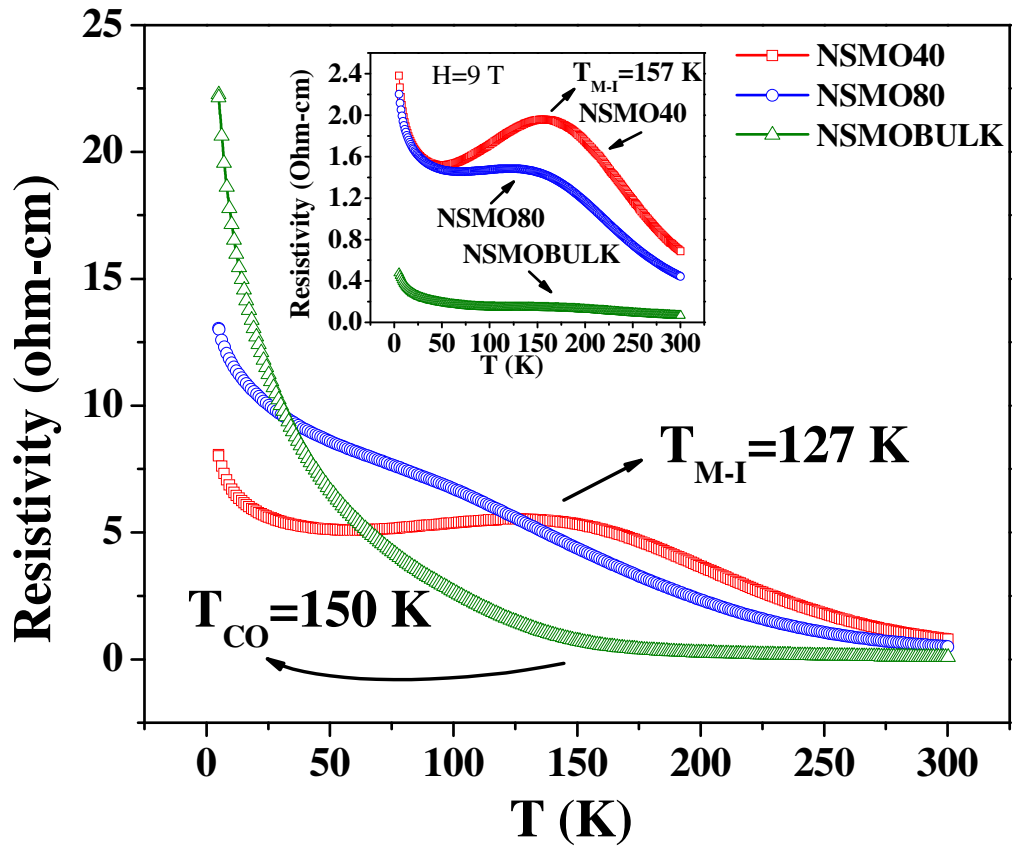

FIG. 4. (Color online) Variation of resistivity of all the three samples as a function of temperature. Inset shows the same under 9 T magnetic field.

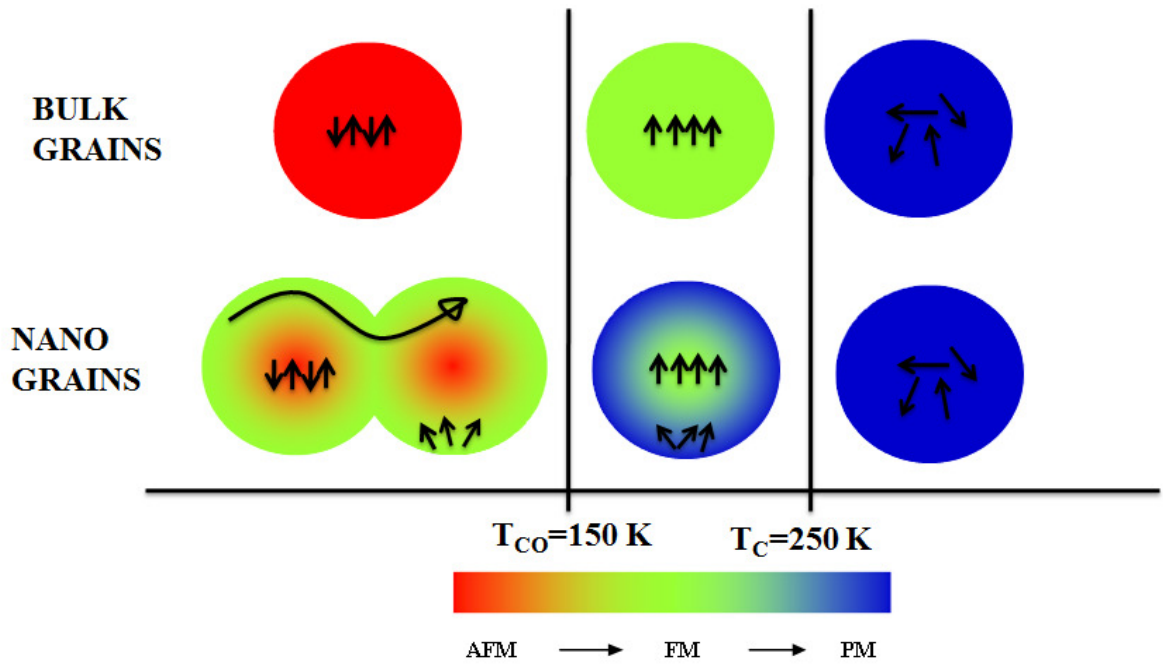

FIG. 5. (Color online) Schematic representation of the phenomenological model for CO/AFM bulk manganites and the corresponding nanograins as the temperature is varied. The curly arrow shows the path for a conduction electron.

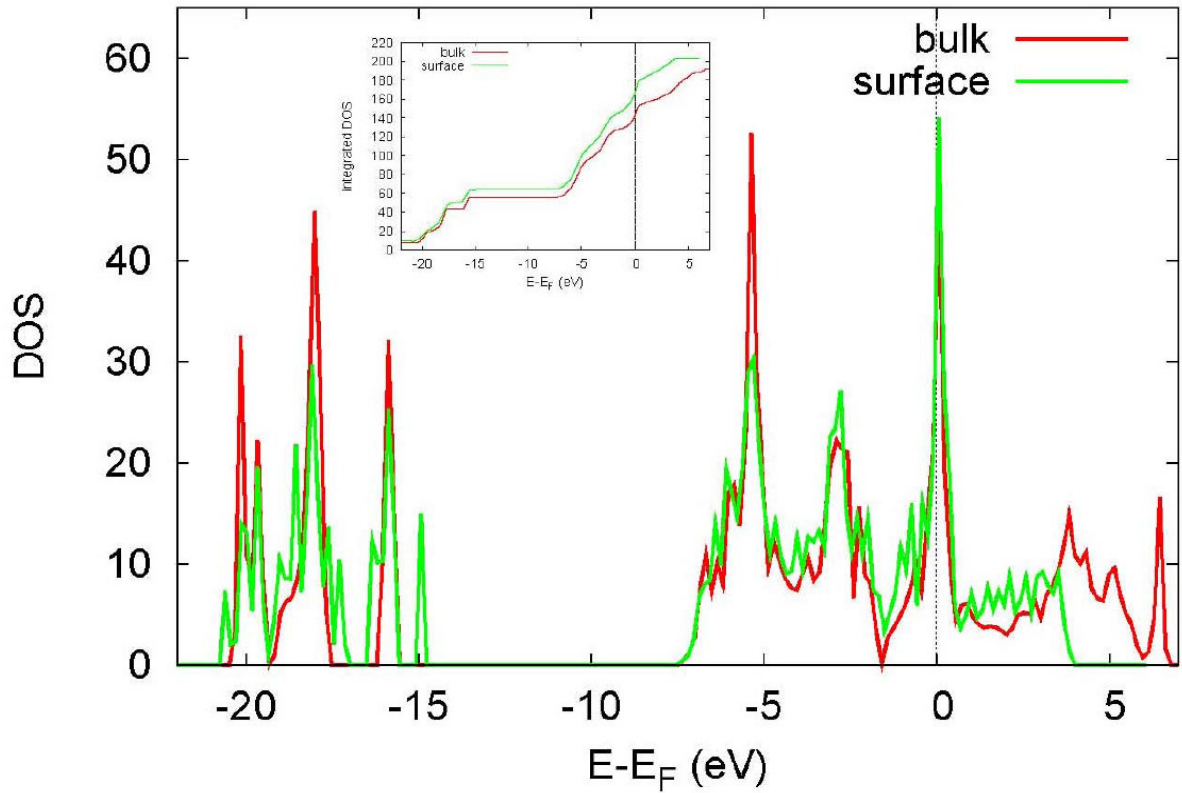

FIG. 6. (Color online) The LDA electronic structure of the bulk and surface $Nd_{0.5}Sr_{0.5}MnO_3$. The Mn-O manifold straddles the Fermi level (E=0) while Nd, Sr derived states extend much below it. The inset represents integrated density of states.

TABLE 1. Lattice parameters, unit cell volume obtained from Rietveld refinement and grains sizes of bulk and nanometric $Nd_{0.5}Sr_{0.5}MnO_3$ samples.

| SAMPLE | a (Å) | b (Å) | c (Å) | V (Å$^3$) | Grain size (nm) |
| --- | --- | --- | --- | --- | --- |
| NSMO40 | 5.4796 | 7.6489 | 5.4619 | 226.773 | 40 |
| NSMO80 | 5.4252 | 7.6413 | 5.4605 | 226.367 | 80 |
| NSMOBULK | 5.4273 | 7.6272 | 5.4701 | 226.439 | ~ 1000 |